\documentclass[12pt]{article}
\usepackage{scicite}
\usepackage{times}
\usepackage{graphicx}
\usepackage{xcolor}
\usepackage{tabularx}

\newenvironment{sciabstract}{%
\begin{quote} \bf}
{\end{quote}}

\newcounter{lastnote}

\title{Quantum state resolved molecular dipolar collisions over four decades of energy}

\author
{Guoqiang Tang$^{1}$, Matthieu Besemer$^{1}$, Stach Kuijpers$^{1}$, Gerrit C. Groenenboom$^{1}$, \\ Ad van der Avoird$^1$, Tijs Karman$^{1\ast\ast}$, Sebastiaan Y.T. van de Meerakker$^1$$^{\ast\ast}$ \\
\normalsize{$^1$Radboud University, Institute for Molecules and Materials}\\
\normalsize{Heijendaalseweg 135, 6525 AJ Nijmegen, the Netherlands}\\
\normalsize{$^{\ast\ast}$To whom correspondence should be addressed;}\\
\normalsize{E-mail: basvdm@science.ru.nl, t.karman@science.ru.nl}}

\date{}

\begin{document}

\baselineskip20pt

\maketitle

\begin{sciabstract}
Collisions between cold polar molecules represent a fascinating research
frontier, but have proven hard to probe experimentally. We report
measurements of inelastic cross sections for collisions between NO and
ND$_3$ molecules at energies between 0.1 and 580 cm$^{-1}$, with full
quantum state resolution. At energies below the $\sim$ 100 cm$^{-1}$
well depth of the interaction potential, we observed backward
glories originating from peculiar U-turn trajectories. At energies below
0.2 cm$^{-1}$, we observed a breakdown of the Langevin capture model,
which we interpreted in terms of a suppressed mutual polarization during
the collision, effectively switching off the molecular dipole moments.
Scattering calculations based on an \emph{ab initio} NO-ND$_3$ potential
energy surface revealed the crucial role of near-degenerate rotational
levels with opposite parity in low-energy dipolar collisions.
\end{sciabstract}%

\newpage
The study of molecular collisions at energies well below 1 kelvin has emerged as a new and exciting research frontier \cite{Dulieu:Book2018}. Interest in cold and ultracold molecular collisions stems from the inherent quantum nature that govern interactions at these low energies, offering a wealth of new opportunities ranging from the production of novel molecular quantum gases and precision measurements in fundamental physics to ultracold chemistry with complete control over all degrees of freedom \cite{Balakrishnan:JCP145:150901,Bohn:Science357:1002,Heazlewood:NatRevChem5:125}. Systems that exhibit electric dipole-dipole interactions are of particular interest, as the long-range and anisotropic dipole-dipole interaction offers an exquisite ``handle'' to control and steer collisions. At sufficiently low temperatures, small external electromagnetic fields can vastly change the interaction, yielding distinctive opportunities to engineer interaction Hamiltonians and to steer collision pathways \cite{Krems:IntRevPhysChem24:99,Sawyer:PCCP13:19059}.

Two main approaches have been developed to experimentally study cold
molecular collisions, and to harvest their prospects for unlocking new
scattering phenomena. In the first, molecules are stored in traps,
yielding long interaction times up to seconds. Spectacular progress has
been reported recently for ultracold collisions between dipolar bialkali
molecules produced through association of ultracold atoms, yielding
temperatures well below 0.1 mK
\cite{Ospelkaus:Science327:853,Ni:Nature464:1324,DeMarco:Science363:853,Matsuda:Science370:1324,Schindewolf:Nature607:677}.
Techniques that use direct slowing or laser cooling of pre-existing
molecules before loading them into a 2D or 3D trap at temperatures in
the 0.1 -- 500~mK range are also coming to the fore
\cite{Meerakker:CR112:4828,Parazzoli:PRL106:193201,Hutzler:CR112:4803,Wu:Science358:645,Segev:Nature572:189,Barry:Nature512:286,Truppe:NatPhys13:1173,Koller:PRL:203401}.
These samples may be cooled further using optical cooling schemes, or by
evaporative or sympathetic cooling schemes if a favorable ratio of
elastic to inelastic collision cross sections is available
\cite{Stuhl:NATURE492:396,Son:Nature580:197}. Molecular collision
dynamics inside the trap is usually inferred from trap loss
measurements, although significant progress has been made recently to
directly probe the collision products
\cite{Hu:Science366:1111,Liu:Nature593:379}.

The second approach relies on crossed molecular beam methods, in which beams of atoms or molecules are made to interact within a well defined crossing volume. Cooling of the molecules of interest into low rotational and translational temperatures occurs within a supersonic expansion by entraining them in a seed gas of noble atoms. Low interaction energies can be achieved by studying intrabeam collisions \cite{Amarasinghe:JPCL8:5153,Gawlas:JPCL11:83,Perreault:Science358:356,Dulitz:JPCA124:3484}, by controlling the beam crossing angle \cite{Amarasinghe:NatChem12:528,Chefdeville:Science341:06092013} or by changing the reagent's speeds using  for instance molecular decelerators \cite{Meerakker:CR112:4828}. Energies as low as 10 mK have been reached using the merged beam technique, in which one of the beams is manipulated to tangentially overlap with the other beam's path \cite{Henson:Science338:234,Gordon:NatChem10:119,Zhelyazkova:PRL125:263401}.

Whereas trapping offers record-low temperatures and presumably the ideal platform to use (ultra)cold molecules in a variety of applications, crossed beam experiments are ideally suited to probe the inherent collision properties with unprecedented detail. One of the most appealing features of the crossed beam technique is the possibility to probe the collisions isolated from any environment, the possibility to tune the collision energy over a wide range, and the availability of a suite of advanced detection techniques to directly measure the collision products state-selectively. Recently, quantum features in the energy dependent state-to-state integral cross sections (ICSs) such as partial wave scattering resonances have been measured at energies down to 0.2 cm$^{-1}$ \cite{DeJongh:Science368:626}. The combination with the velocity map imaging (VMI) technique allowed for measurements of differential cross sections (DCSs) that directly revealed the quantum waves underlying the scattering event \cite{Margulis:NatCom11:3553,DeJongh:NatChem14:538}.

Despite this vast progress, there are still many outstanding questions in our understanding of cold molecular collisions. Depending on the collision energy, different theoretical approaches are generally followed to describe the collision process. At sufficiently low energies, the collisions may be described by universal laws and capture theory \cite{Bohn:NJP11:055039,Cavagnero:NJP11:055040}, based on interaction potentials that mainly contain long-range electrostatic interactions \cite{Avdeenkov:PRL90:043006}. At higher energies, on the other hand, full quantum scattering calculations based on accurate \emph{ab initio} potential energy surfaces (PESs) are required. It is at present unclear how and at which energies these approaches connect to each other. In addition, at low energies the collisions become extremely sensitive to the exact topology of the PES; a small modification of interaction potentials can change cross sections by orders of magnitude \cite{Frye:NJP17:045019}. This fact has major implications for the success of evaporative and sympathetic cooling schemes, as these schemes strongly rely on the exact collision properties of the system across a relatively large range of energies. As a result, theoretical models urgently need experimental data to become predictive. Last but not least, in describing cold collisions between polar molecules, it is important to realize that polar molecules in isolated quantum states formally do not possess a permanent dipole moment. Effective dipole moments must be induced, for instance by an external electric field or the proximity of another polar molecule, but it is at present unclear how and under which conditions this induced dipole moment starts affecting the scattering behavior.    

It has proven extremely difficult to experimentally probe low-energy molecular collisions in the required detail, in particular for collisions between two dipolar molecules. Ideally, one would like to experimentally track how two molecules approach each other as a function of the collision energy, interact through their evolving potential energy landscapes, and reemerge as final products \cite{Bohn:Science357:1002}. Only then can a complete picture of collision events be obtained, from start to finish and from hot to cold, and validate various theoretical models along the way. Although all methodological ingredients to achieve this result have in principle been developed, the combination of all techniques in a single experiment has been deemed impedingly difficult. State-to-state crossed beam experiments below $\approx$ 1~kelvin have thus far only been realized for experiments in which a molecule of interest scatters with beams of He atoms or H$_2$ molecules, making use of the high densities available in neat beams and the low mass of the collision partner to decrease the reduced mass \cite{DeJongh:Science368:626,Chefdeville:Science341:06092013}. Merged beams have been successful in reaching lower energies \cite{Henson:Science338:234}, but its application to state-selective bi-molecular systems has remained elusive. A state-to-state collision experiment using two polar molecules has been performed for the OH-NO system, but the low densities resulting from the necessity of using two seeded molecular beams has hampered the extension of these experiments to energies below $\approx$ 100 K \cite{Kirste:Sience338:1060}.

Here we present a joint experimental and theoretical study of inelastic
collisions between NO ($X\,^2\Pi_{1/2}, v=0, j=1/2f$ 
referred to hereafter as NO($1/2f$)) radicals and ND$_3$ ($X\, ^1
A_1^{\prime}, v_2=0, j_k^p=1_1^-$, referred to hereafter
as ND$_3$($1_1^-$)) molecules. We used the combination of Stark
deceleration of NO, hexapole state selection of ND$_3$ and VMI detection
in both crossed and merged beam arrangements to measure fully
state-resolved integral and differential cross sections at collision
energies spanning nearly four orders of magnitude between 0.1~cm$^{-1}$
and 580~cm$^{-1}$. This energy range encompasses the temperature scales
relevant to atmospheric processes and combustion ($T\geq 100$~K),
astrochemistry ($T\approx 1$~K) and approaches the ultracold pure
quantum regime ($T \leq 1$~mK). The scattering results revealed three
distinctive energy regimes with different mechanisms governing the
collisions. At the highest energies, between 100~and 580~cm$^{-1}$, we
observed correlated NO - ND$_3$ excitations, which could be understood
from the electrostatic multipole contributions to the interaction
potential. In the intermediate energy regime, between 10~and
100~cm$^{-1}$, the collision energy became similar in magnitude to the
well depth of the angularly averaged potential. In this region we
observed a peculiar scattering mechanism in which the collision partners
experience half orbiting or U-turn-like trajectories, resulting in a
prominent backward component in the scattering distribution. At the
lowest energies, we found an intriguing dependence of the ICS on
collision energy, which was interpreted in terms of a breakdown of the
Langevin capture model when the dipole-dipole interaction energy becomes
similar to the energy splittings between near-degenerate rotational
levels with opposite parity, effectively switching off the molecular
dipole moments. Excellent agreement was obtained with the cross sections
derived from quantum coupled-channels scattering calculations based on
a new \emph{ab initio} NO-ND$_3$ PES throughout the entire energy range.\\

{\bf Crossed and merged beam scattering}\\

\begin{figure}[!htb]
	\centering
	\includegraphics[width=\linewidth]{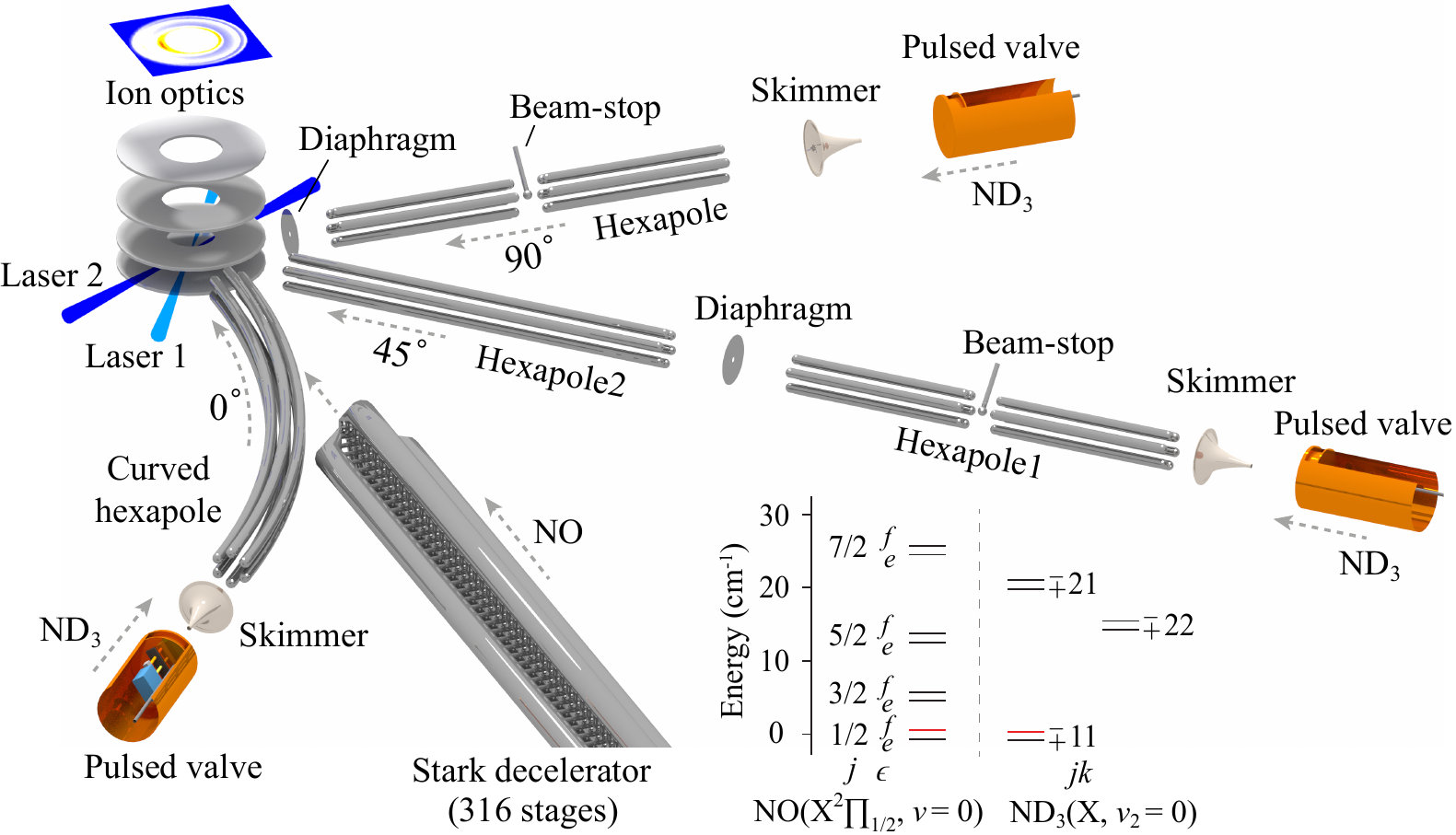}
	\caption{Schematic representation of the experimental setup and the rotational energy level scheme of
the NO radical and para-ND$_3$. A state-selected and velocity tunable beam of NO radicals is produced
using a 2.6-meter-long Stark decelerator, and crossed with a hexapole state-selected beam of ND$_3$ molecules using either a 0$^\circ$, 45$^\circ$ or 90$^\circ$ angle of incidence. Only the last section of the decelerator is shown. Scattered NO radicals are detected state-selectively using velocity map imaging.
The rotational levels of NO and ND$_3$ are split into two components of opposite parity. The energy splittings are largely exaggerated for reasons of clarity. The initial levels selected by the Stark decelerator and hexapoles are indicated in red.}	
	\label{fig:setup}
\end{figure}

We used a crossed molecular beam apparatus that is schematically shown in Fig. \ref{fig:setup} and explained in more detail in the Supporting Materials (SM). Packets of NO($1/2f$) radicals with a tunable velocity were produced by passing a beam of $\approx$ 5\% NO seeded in either krypton, xenon or argon through a 2.6~meter long Stark decelerator. Packets of state-selected ND$_3$($1_1^-$) molecules were produced by passing a beam of $\approx$ 5\% ND$_3$ seeded in rare gas atoms through electrostatic hexapoles. To cover a large range of collision energies, we used assemblies of straight hexapoles at 45 and 90 degree intersection angle with the Stark decelerator axis, as well as a curved hexapole to merge the ND$_3$ and NO packets at near zero degree intersection angle. The relatively small dipole moment of 0.16~D for NO compared to the 1.5~D moment for ND$_3$ ensured that both beams could be merged without significantly affecting the NO packet (SM). The straight hexapole beamlines featured a beamstop and diaphragm located such that the rare gas atoms were effectively eliminated from the ND$_3$ molecular beam pulse. The scattered NO radicals were state-selectively detected using a two-color laser ionization scheme and velocity mapped on a two dimensional detector (SM).\\

{\bf Velocity mapped ion images probing differential cross sections}\\

\begin{figure}[!htb]
	\centering
	\includegraphics[width=\linewidth]{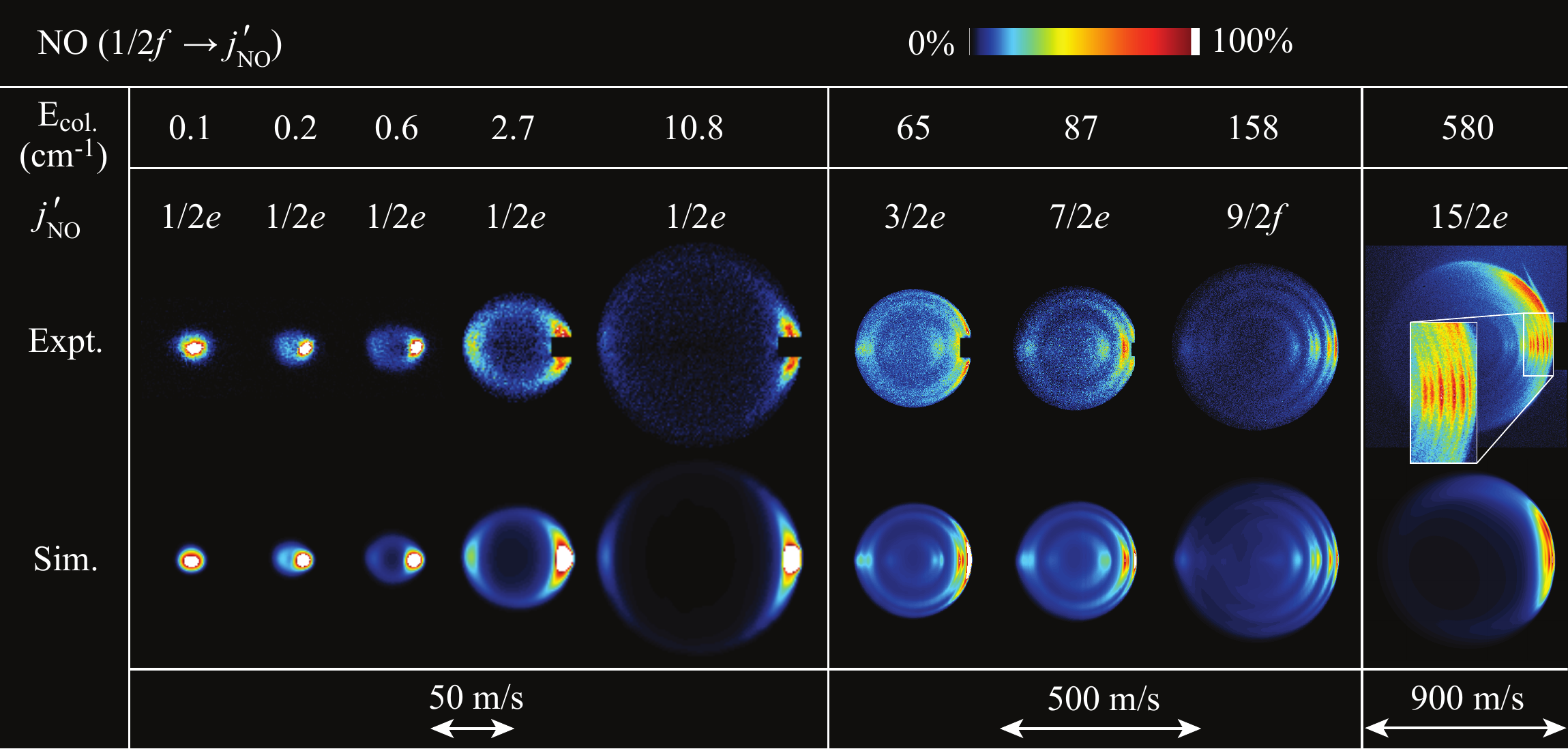}
	\caption{Selection of the experimental (Expt) and simulated (Sim) velocity map ion images for the scattering process NO ($1/2f$) + ND$_3$ ($1_1^-$) $\rightarrow$ NO ($j^{\prime}_{\mathrm{NO}}$) + ND$_3$ ($j^{\prime}_k$) as a function of collision energy. The images are presented such that the relative velocity vector is oriented horizontally, with the forward direction on the right side of the image. Small
segments of the images around forward scattering are masked due to imperfect state selection of the NO packet. The final state $j^{\prime}_{\mathrm{NO}}$ of NO probed by the lasers is indicated above each image. Note the three different velocity scales used to cover the large range of collision energies.}	
	\label{fig:images}
\end{figure}

We measured velocity mapped ion images for various final NO states at 21 different collision energies $E_\mathrm{col}$ (see SM for all images and Fig. \ref{fig:images} for a selection), directly probing the DCSs of the scattering process. At high energies, multiple concentric rings were observed that correspond to the coincident excitation of ND$_3$ into excited $j_k$ levels. By virtue of the Stark decelerator, the resolution in most images was sufficient to probe individual  transitions up to $\Delta j=8$ and $\Delta k=6$, however, excitation to near-degenerate parity components within a given ND$_3$($j_k$) level could not be resolved. As the energy was reduced, the images became smaller and featured fewer rings.  At sufficiently low energies, excitation of ND$_3$ became energetically inaccessible, and only a single ring was observed that converged to a single point at the lowest probed collision energy of 0.1 cm$^{-1}$. At all energies, most scattering was observed in the forward direction, with a striking and rather intense narrow backscattered feature appearing for collision energies below $\approx$ 100 cm$^{-1}$.

We compared the experimental images to simulated images based on
coupled-channels (CC) scattering calculations of state-to-state cross
sections using our \emph{ab initio} NO-ND$_3$ PES and the kinematics of
the experiment (SM). Unfortunately, the high density of rotational
levels in both NO and ND$_3$, in combination with the dipole-dipole
interaction extending to large radial distances $R$, made full CC
calculations computer intensive. To limit the
computational cost we truncated the channel basis, depending on
the collision energy (SM). The open-shell character and
$\Lambda$-doublet splittings of NO were taken into account as these factors were
probed in the experiment, but the inversion splittings in ND$_3$ were
neglected as these splittings were not experimentally resolved. For the energies
158 and 580 cm$^{-1}$, the NO radical was treated as a closed-shell
molecule, which further reduced the size of the channel basis. The
validity of this approximation at high energies was verified by
comparing the cross sections at 65 cm$^{-1}$ obtained from the
closed-shell and open-shell calculations (SM).

The angular and radial distributions derived from the experimental and
simulated images showed excellent agreement (SM), validating theory
to uncover the mechanisms underlying these observations and the role of
the dipole-dipole interaction in them. Not surprisingly, our
calculations revealed that the scattering dynamics is dominated by
rather different mechanisms across the four orders of magnitude wide
collision energy range probed in the experiments. Trends in the energy
dependence of the pair-correlated NO-ND$_3$ excitations could be related
to the electrostatic multipole contributions to the interaction
potential. The
dipole-dipole and quadrupole-dipole interaction dominate the NO-ND$_3$
attractive forces at long range and govern ND$_3$ excitations with
$\Delta k=0$, and these transitions were found strongest at energies
above $\sim 60$ cm$^{-1}$. At lower energies, transitions with $|\Delta
k| =3$ gained intensity, which require weaker and more short-ranged
terms involving the octupole moment of ND$_3$. This counter-intuitive
result originated from a balance between the energy gap law that favored
transitions with low excitation energy, and the relative strength of the
multipole interactions (SM). Model calculations showed that if the final states with $\Delta k=0$ and $|\Delta
k| =3$ were degenerate, excitations with $\Delta k=0$ were then indeed preferred throughout, confirming the role of the energy gap (SM). \\

{\bf Backward glories}\\

\begin{figure}[!htb]
	\centering
	\includegraphics[width=\linewidth]{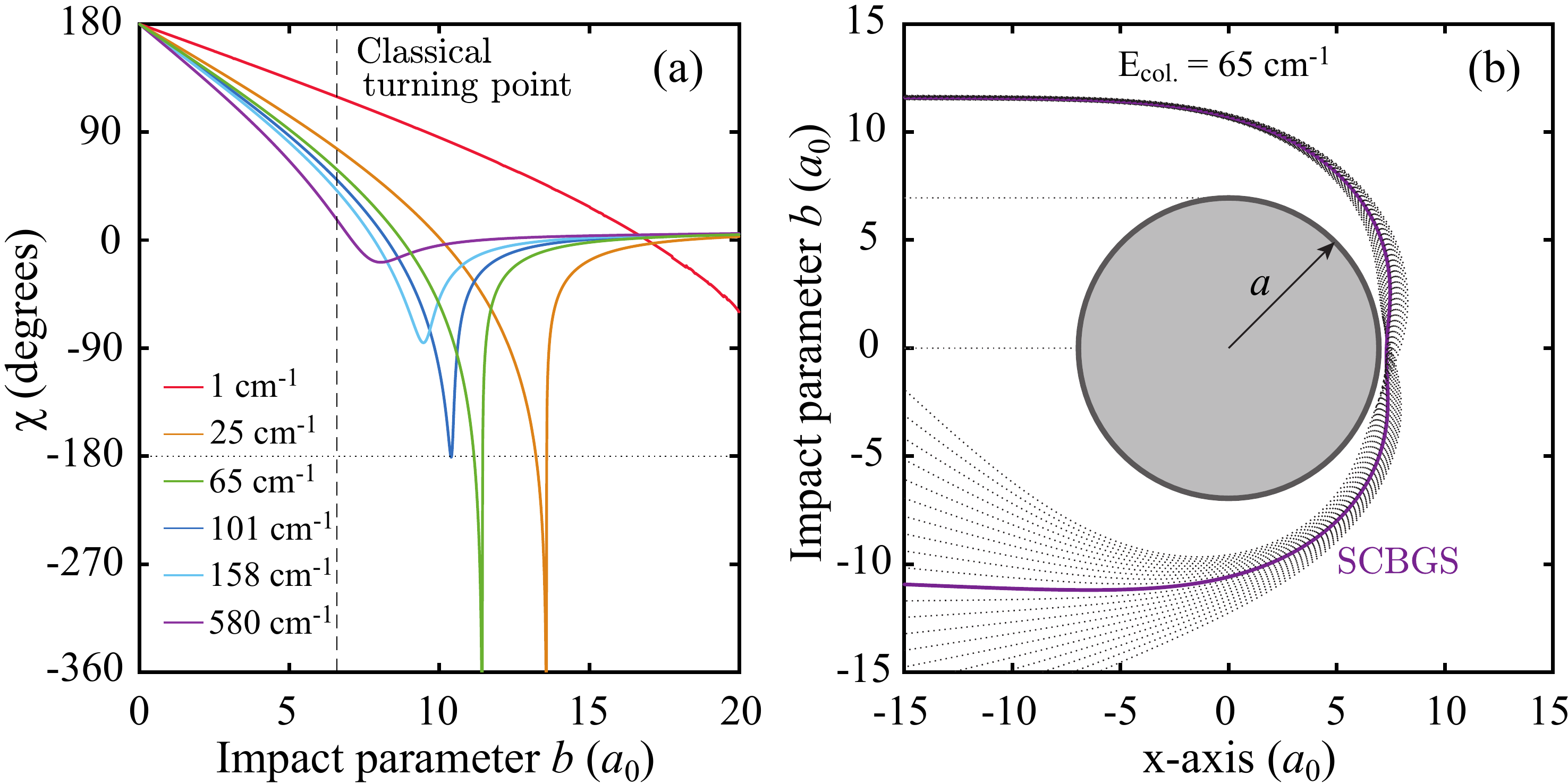}
	\caption{Illustration of the origin of backward glories. (a) Deflection $\chi$ as a function of the impact parameter $b$ resulting from classical trajectory calculations for elastic collisions between NO and ND$_3$ for selected collision energies. At energies below 101~cm$^{-1}$, backscattering ($\chi$ = -180$^\circ$) can occur for impact parameters well above the classical turning point. (b) Swarm of classical trajectries for a narrow range of impact parameters above the classical turning point leading to scattering in the backward direction at a collision energy of 65 cm$^{-1}$. The Soft Collision Backward Glory Scattering (SCBGS) trajectory leading to perfect backscattering $\chi=-180^{\circ}$ is indicated in purple.}	
	\label{fig:glories}
\end{figure}

The dominant forward scattering observed in almost all images was found
to originate from the dipole-dipole interaction, and disappeared when
this term was switched off in model calculations (SM). By contrast,
the narrow backscattered feature observed below collision energies of
$\approx$ 100 cm$^{-1}$ persisted, even in the absence of the
dipole-dipole interaction. Using classical trajectory calculations, we found 
that this feature originated from
peculiar scattering trajectories with impact parameters within a
surprisingly narrow range above the classical turning point (see Fig.
\ref{fig:glories}(b)). Common wisdom predicts that such soft or glancing
collisions result in forward or side scattering, yet, we observed
half-orbiting trajectories that swung around the scattering partner to
exit at a near-exact backscattered angle of $\chi= -180^{\circ}$ (we used the common convention of the deflection angle $\chi$ which is related to the observable scattering angle $\theta=|\chi|$). We
found that this peculiar behavior was related to the well-known
phenomenon of rainbow scattering, resulting in a maximum deflection
(rainbow) angle that occured for trajectories with a specific (rainbow)
impact parameter. We found that for energies below the well depth of the
angularly averaged interaction potential (101.1 cm$^{-1}$ for
NO-ND$_3$), the attraction could become sufficiently strong to push the
rainbow angle to $\chi=-180^{\circ}$, defining the onset of orbiting (Fig. \ref{fig:glories}(a)). This fact resulted in a strong enhancement of the
DCS (that in a classical treatment diverges as $1/\sin\theta$) at
$\theta=180^{\circ}$, similar to the more familiar glory effect found
in forward scattering. Such glories do not represent a large fraction of the scattering, but they accumulate intensity in a small region.  The `soft-collision backward glories' as found
here are known in optics \cite{Langley:AO30:3459},
but their observation in scattering has thus far been scarce and limited
to high energy collisions of alkali systems that have deep
potential wells \cite{Duren:CPL64:357,Loesch:JCP99:9598}. Yet, they
should be obiquitous in collisions between dipolar molecules
characterized by only modest well depths provided that the collision
energy is sufficiently low.\\

{\bf Local maximum in low-energy integral cross sections}\\

\begin{figure}[!htb]
	\centering
	\includegraphics[width=\linewidth]{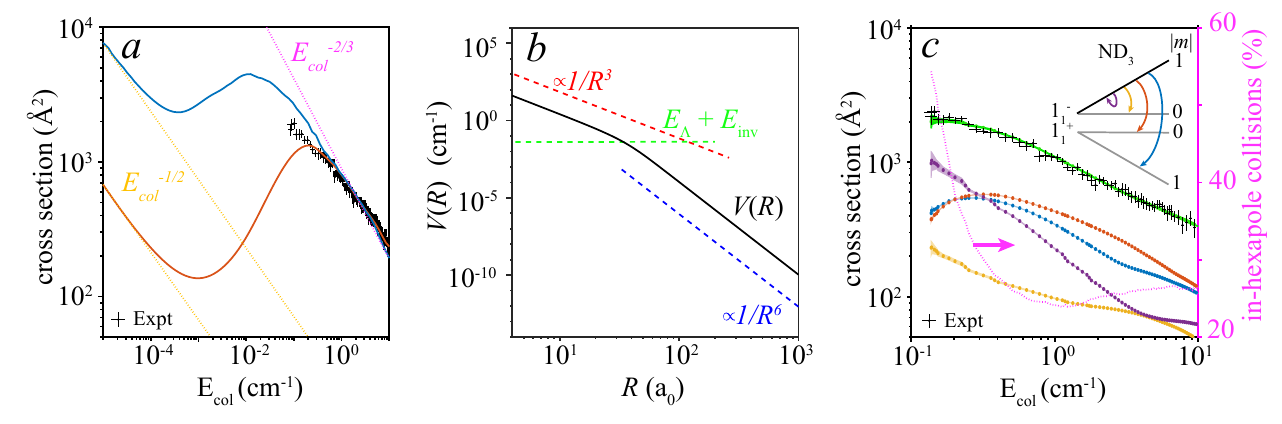}
	\caption{Low-energy scattering behavior. (a) Experimentally observed ICS for the scattering process NO ($1/2f$) + ND$_3$ ($1_1^-$) $\rightarrow$ NO ($1/2e$) + ND$_3$ ($j^{\prime}_k$) as a function of collision energy together with the theoretically predicted ICS without (blue curve) and with (red curve) taking the inversion doublet structure of ND$_3$ into account. Data is accumulated using a continuous cycle over collision energies. Vertical error bars represent statistical uncertainties at 95\% of the confidence interval (all panels). Horizontal error bars represent uncertainties in energy calibration (all panels). Scattering signals were corrected for flux-to-density effects taking the spatial, temporal and velocity distributions of both the NO and ND$_3$ packets into account, assuming that detectable collisions only occur when the merging process has completed (see SM). Experimental cross sections are scaled to the theoretical ones. (b) The NO-ND$_3$ interaction strength as a function of the radial distance $R$ between the molecules. The sum of the $\Lambda$-doublet splitting of NO ($E_{\Lambda}$) and the inversion splitting of ND$_3$ ($E_\mathrm{inv}$) is indicated by the green dashed line. (c) As in (a), but taking the partial beam overlap inside the curved hexapole and the $m$-resolved electric-field dependent ICSs into account. The cross sections resulting from numerical trajectory simulations for the elastic and inelastic channels in ND$_3$ that lead to detectable NO (1/2\emph{e}) products are indicated (colored data points). The statistical errors associated with numerical trajectory simulations are indicated by the colored areas. Experimental cross sections are scaled to the sum of the individual channels (green). The percentage of collisions that take place inside the curved hexapole is indicated (magenta dashed line, right axis).}	
	\label{fig:ICS}
\end{figure}

At the lowest energies it was difficult to discern structure in the
small images. We therefore made complementary ICS measurements for
collisions that de-excite NO radicals from the $1/2f$ to the $1/2e$
level by scanning the collision energy from 0.1 to 10 cm$^{-1}$. The
collision signals were converted into relative cross sections taking the
spatial and temporal overlap of both beams into account, assuming that
detectable collision products could only emerge after the curved hexapole
when the merging process had completed (SM). We observed a rapidly
increasing cross section as the energy was reduced (Fig.
\ref{fig:ICS}a). At low energies, one may expect the cross sections to
follow the Langevin capture model, which for a dipole-dipole interaction
with long-range $1/R^3$ dependence predicts a cross section that scales
with $E_\mathrm{col}^{-2/3}$. We found that the cross section indeed
followed this energy dependence, but started to deviate for energies below
$\sim 1$ cm$^{-1}$.

The cross section that followed from CC calculations reproduced the
Langevin cross section at high energies, and furthermore revealed an
intricate dependence on collision energy and a large sensitivity to the
exact parameters used in the calculations \cite{Bohn:PRA63:052714}. We
first used the CC calculations described above that take the
$E_{\Lambda}=0.0119$ cm$^{-1}$ $\Lambda$-doublet splitting between the
$1/2f$ and $1/2e$ states of NO into
account but that neglected the $E_\mathrm{inv}= 0.053$ cm$^{-1}$ inversion
splitting between the
$1_1^-$ and $1_1^+$ states of ND$_3$, and calculated the ICS for energies
extending into the ultracold regime (blue curve in Fig. \ref{fig:ICS}a).
Excellent agreement with the experiment was obtained, except for
energies below 0.2 cm$^{-1}$ where we found that the predicted ICS could
change by orders of magnitude by manually adjusting the value for
$E_{\Lambda}$ between zero and 0.3 cm$^{-1}$ (SM). Clearly, the
approximation to neglect $E_\mathrm{inv}$ in the calculations was no
longer valid at low energies, and we performed CC calculations that
included $E_\mathrm{inv}$ at the expense of a smaller basis set that was
appropriate at these energies (red curve in Fig. \ref{fig:ICS}a). Both
curves nearly coincided at high energies and converged to the Langevin
cross section, but at energies below 0.2 cm$^{-1}$ they showed a local maximum
(LM) that peaks at different energies before they entered the Wigner
regime where the cross section scaled with $E_\mathrm{col}^{-1/2}$
following the Wigner threshold law.

To understand this intricate LM in the ICS, one should realize
that the initial states of NO and ND$_3$ in the experiments ---the $1/2f$
and $1_1^-$ state, respectively--- do not have permanent dipole moments.
Effective dipole moments are induced by electric fields, but also by
dipole-dipole interactions, by mixing the $1/2f$ and $1/2e$ states of NO
and the $1_1^-$ and $1_1^+$ states of ND$_3$. Figure~\ref{fig:ICS}b
shows the effective interaction energy $V(R)$ as seen by the approaching
NO and ND$_3$ molecules calculated as a function of $R$. We used a
simple model with two two-level systems that accounts only for two
opposite parity states of each molecule, at energies $\pm E_{\Lambda}/2$
and $\pm E_\mathrm{inv}/2$, and the dipole-dipole interaction that
couples collision channels where both molecules change their parity,
yielding off-diagonal elements that scale with $\mu_{NO} \mu_{ND_3}/R^3$
(SM). At short distances, typically probed at high energies, the
dipole-dipole interaction is strong compared to the inversion
splittings, and mixes the opposite parity states of both molecules. This
mutual polarization leads to $1/R^3$ dipolar interactions also without
applying an external field to polarize the molecules
\cite{Bohn:PRA63:052714,Cavagnero:NJP11:055040}, which
results in the observed $E_\mathrm{col}^{-2/3}$ Langevin capture. As the
collision energy is reduced, this cross section increases reflecting a
larger capture impact parameter. At larger distances, however, the
interaction potential begins to deviate from $1/R^3$ and switches to a
$1/R^6$ dependence when the dipole-dipole interaction becomes weaker
than the parity splittings, and can no longer polarize the molecules.
This transition occurs where the dipole-dipole interaction energy is
comparable to the sum of the energy splittings between levels of
opposite parity in both molecules, explaining the dependence on the
inclusion of $E_\mathrm{inv}$. Hence, we observe a breakdown of the
Langevin capture model, leading to the characteristic LM in the cross
section. The field-free dipole-dipole capture we observe towards higher
energies may also be interpreted in terms of nonadiabatic transitions as
described in
Ref.~\cite{Troe:JPCA114:9762,Troe:JPCA115:5027,Troe:JCP130:014304}.

The experimentally observed scattering behavior deviated significantly
from the theoretical prediction below 0.2 cm$^{-1}$, and did not show the local maximum. In this region, we
found that the cross section not only depended critically on
$E_{\Lambda}$ and $E_\mathrm{inv}$, but also responded extremely
sensitively to external electric fields \cite{Kajita:EPJD38:315}. Even
fields of only 100 - 1000 V/cm caused the cross section to increase by an
order of magnitude, even though such fields hardly changed the energy
splittings. By contrast, they broke the parity selection rule of the
dipole-dipole interaction, resulting in a large cross section increase
for collisions in which NO was transferred to the 1/2\emph{e} state but
ND$_3$ remained in the $1_1^-$ state (SM). At the lowest collision
energies accessible to our experiment, the NO and ND$_3$ molecules
travelled with near-matching velocities causing a partial overlap of both
beams already in the final section of the curved hexapole where large
inhomogeneous electric fields were present (SM). Although high-field
seeking, simulations showed that NO (1/2\emph{e}) collision products were
not significantly influenced by the hexapole field, and had a high
probability to be detected by the laser further downstream. In addition
to changing the cross sections, these in-hexapole collisions also
increased the effective beam intersection angle and collision energy for
given mean velocities of the NO and ND$_3$ packets. To account for both
effects, we calculated the electric field dependent ICSs for all
possible $m$-resolved transitions in ND$_3$ (SM), and simulated for
each channel the expected cross section to produce detectable NO
(1/2\emph{e}) collision products as a function of the collision energy
probed in the experiment (colored curves in Fig. \ref{fig:ICS}c). We
found excellent agreement between the experiment and theoretical
prediction when including the in-hexapole collisions in our analysis
(green curve in Fig. \ref{fig:ICS}c), and could interpret the slowed rate of increase in cross section observed at the lowest energies as the onset of the local maximum. We found that at the lowest
energies up to 50\% of the detectable collision events occured inside the
strong field of the hexapole where the elastic ND$_3$ channel started
dominating the collision dynamics. The partial overlap inside the curved
hexapole is inherent to the merged beam approach with matching
velocities, but may in future experiments be circumvented by tagging one
of the beams using an optical excitation pulse when the merging is
completed, effectively setting a time zero for the beam overlap. This would also reduce the minimum collision energy achievable, which would then solely be determined by the velocity spreads of both packets.  

The collision behavior including the LM found here should be ubiquitous in the scattering of two polar molecules with near-degenerate energy levels of opposite parity. It is exactly this kind of species that is most amenable to Stark deceleration, hexapole state selection and electrostatic trapping alike \cite{Meerakker:CR112:4828}. The energy $E_\mathrm{LM}$ at which the LM peaks, as well the cross section $\sigma_\mathrm{LM}$ found at this energy, can be estimated from capture theory (SM) and are given for a selection of relevant systems in Table \ref{tab:Esp_Ecol} (see SM for an extended table). The value for $E_\mathrm{LM}$ exclusively depends on the sum of the parity splitting energies of both colliders, shifting the LM to higher energies for systems with larger splittings.\\

{\bf Concluding remarks}\\

The success attained here to measure completely state-resolved ICSs and DCSs for polar bimolecular collisions at energies tunable by nearly four orders of magnitude -- down to 0.1 cm$^{-1}$ -- reveals detailed views on collisions between cold polar molecules, opens new vistas to control the collision dynamics using external fields, and offers new insights to assess the feasibility of future experiments. The existence of backward glories in the DCSs, field-free dipolar collisions by mutual polarization, the breakdown of the capture model, and the extreme sensitivity to external fields at low energies, all provide extraordinary opportunities to control the collision dynamics. The collision energies we achieved are lower than the typical interaction energy of a polar molecule with external fields, yielding distinctive opportunities to manipulate the dipole-dipole interaction. Tailored external electric fields could be applied in a dedicated section after the curved hexapole, in which both the strength and direction of the field could be adjusted. As the molecule's dipole moment is either oriented along ($\uparrow$) or against ($\downarrow$) the electric field direction depending on the quantum state, the interaction could be effectively switched from repulsive ($\uparrow\uparrow$) to attractive ($\uparrow\downarrow$), tuned in strength by controlling the degree of polarization, or controlled by resonances associated with field-linked bound states~\cite{Avdeenkov:PRL90:043006}.
Microwave fields could be applied to dress the rotational levels, which could enhance or supress collision channels \cite{Avdeenkov:PRA86:022707,Karman:PRL121:163401,Lassabliere:PRL121:163402,Anderegg:Science373:779,Yan:PRL125:063401}.
Collision induced recoil energies from Stark-shifted energy levels could directly be recorded using VMI, yielding pair-correlated $m$-resolved collision cross sections revealing the relative orientation of the products. Application of various such external control knobs to tune and follow the dynamics in collisions between cold polar molecules, a Holy Grail in the field for two decades, are now within reach.


\begin{thebibliography}{10}

\bibitem{Dulieu:Book2018}
O.~Dulieu, A.~Osterwalder, eds., {\it Cold Chemistry\/} (The Royal Society of
  Chemistry, 2018).

\bibitem{Balakrishnan:JCP145:150901}
N.~Balakrishnan, Perspective: Ultracold molecules and the dawn of cold
  controlled chemistry, {\it J. Chem. Phys.\/} {\bf 145},   150901 (2016).

\bibitem{Bohn:Science357:1002}
J.~L. Bohn, A.~M. Rey, J.~Ye, Cold molecules: Progress in quantum engineering
  of chemistry and quantum matter, {\it Science\/} {\bf 357},   1002--1010
  (2017).

\bibitem{Heazlewood:NatRevChem5:125}
B.~R. Heazlewood, T.~P. Softley, Towards chemistry at asolute zero, {\it Nat.
  Rev. Chem.\/} {\bf 5},   125--140 (2021).

\bibitem{Krems:IntRevPhysChem24:99}
R.~V. Krems, Molecules near absolute zero and external field control of atomic
  and molecular dynamics, {\it Int. Rev. Phys. Chem.\/} {\bf 24},   99--118
  (2005).

\bibitem{Sawyer:PCCP13:19059}
B.~C. Sawyer, {\it et~al.\/}, Cold heteromolecular dipolar collisions, {\it
  Phys. Chem. Chem. Phys.\/} {\bf 13},   19059--19066 (2011).

\bibitem{Ospelkaus:Science327:853}
S.~Ospelkaus, {\it et~al.\/}, Quantum-state controlled chemical reactions of
  ultracold potassium-rubidium molecules, {\it Science\/} {\bf 327},   853--857
  (2010).

\bibitem{Ni:Nature464:1324}
K.~K. Ni, {\it et~al.\/}, Dipolar collisions of polar molecules in the quantum
  regime, {\it Nature\/} {\bf 464},   1324--1328 (2010).

\bibitem{DeMarco:Science363:853}
L.~De~Marco, {\it et~al.\/}, A degenerate fermi gas of polar molecules, {\it
  Science\/} {\bf 363},   853--856 (2019).

\bibitem{Matsuda:Science370:1324}
K.~Matsuda, {\it et~al.\/}, Resonant collisional shielding of reactive
  molecules using electric fields, {\it Science\/} {\bf 370},   1324--1327
  (2020).

\bibitem{Schindewolf:Nature607:677}
A.~Schindewolf, {\it et~al.\/}, Evaporation of microwave-shielded polar
  molecules to quantum degeneracy, {\it Nature\/} {\bf 607},   677--681 (2022).

\bibitem{Meerakker:CR112:4828}
S.~Y.~T. van~de Meerakker, H.~L. Bethlem, N.~Vanhaecke, G.~Meijer, Manipulation
  and control of molecular beams, {\it Chem. Rev.\/} {\bf 112},   4828--4878
  (2012).

\bibitem{Parazzoli:PRL106:193201}
L.~P. Parazzoli, N.~J. Fitch, P.~S. \ifmmode~\dot{Z}\else \.{Z}\fi{}uchowski,
  J.~M. Hutson, H.~J. Lewandowski, Large effects of electric fields on
  atom-molecule collisions at millikelvin temperatures, {\it Phys. Rev.
  Lett.\/} {\bf 106},   193201 (2011).

\bibitem{Hutzler:CR112:4803}
N.~R. Hutzler, H.-I. Lu, J.~M. Doyle, The buffer gas beam: An intense, cold,
  and slow source for atoms and molecules, {\it Chem. Rev.\/} {\bf 112},
  4803--4827 (2012).

\bibitem{Wu:Science358:645}
X.~Wu, {\it et~al.\/}, A cryofuge for cold-collision experiments with slow
  polar molecules, {\it Science\/} {\bf 358},   645--648 (2017).

\bibitem{Segev:Nature572:189}
Y.~Segev, {\it et~al.\/}, Collisions between cold molecules in a
  superconducting magnetic trap, {\it Nature\/} {\bf 572},   189--193 (2019).

\bibitem{Barry:Nature512:286}
J.~F. Barry, D.~J. McCarron, E.~B. Norrgard, M.~H. Steinecker, D.~DeMille,
  Magneto-optical trapping of a diatomic molecule, {\it Nature\/} {\bf 512},
  286--289 (2014).

\bibitem{Truppe:NatPhys13:1173}
S.~Truppe, {\it et~al.\/}, Molecules cooled below the doppler limit, {\it Nat.
  Phys.\/} {\bf 13},   1173--1176 (2017).

\bibitem{Koller:PRL:203401}
M.~Koller, {\it et~al.\/}, Electric-field-controlled cold dipolar collisions
  between trapped {CH}$_3${F} molecules, {\it Phys. Rev. Lett.\/} {\bf 128},
  203401 (2022).

\bibitem{Stuhl:NATURE492:396}
B.~K. Stuhl, {\it et~al.\/}, Evaporative cooling of the dipolar hydroxyl
  radical, {\it Nature\/} {\bf 492},   396--400 (2012).

\bibitem{Son:Nature580:197}
H.~Son, J.~J. Park, W.~Ketterle, A.~O. Jamison, Collisional cooling of
  ultracold molecules, {\it Nature\/} {\bf 580},   197--200 (2020).

\bibitem{Hu:Science366:1111}
M.-G. Hu, {\it et~al.\/}, Direct observation of bimolecular reactions of
  ultracold {KR}b molecules, {\it Science\/} {\bf 366},   1111--1115 (2019).

\bibitem{Liu:Nature593:379}
Y.~Liu, {\it et~al.\/}, Precision test of statistical dynamics with
  state-to-state ultracold chemistry, {\it Nature\/} {\bf 593},   379--384
  (2021).

\bibitem{Amarasinghe:JPCL8:5153}
C.~Amarasinghe, A.~G. Suits, Intrabeam scattering for ultracold collisions,
  {\it J. Phys. Chem. Lett.\/} {\bf 8},   5153--5159 (2017).

\bibitem{Gawlas:JPCL11:83}
K.~Gawlas, S.~D. Hogan, Rydberg-state-resolved resonant energy transfer in cold
  electric-field-controlled intrabeam collisions of {NH}$_3$ with {R}ydberg
  {H}e atoms, {\it J. Phys. Chem. Lett.\/} {\bf 11},   83--87 (2020).

\bibitem{Perreault:Science358:356}
W.~E. Perreault, N.~Mukherjee, R.~N. Zare, Quantum control of molecular
  collisions at 1 kelvin, {\it Science\/} {\bf 358},   356--359 (2017).

\bibitem{Dulitz:JPCA124:3484}
K.~Dulitz, M.~van~den Beld-Serrano, F.~Stienkemeier, Single-source, collinear
  merged-beam experiment for the study of reactive neutral–neutral
  collisions, {\it J. Phys. Chem. A\/} {\bf 124},   3484--3493 (2020).

\bibitem{Amarasinghe:NatChem12:528}
C.~Amarasinghe, {\it et~al.\/}, State-to-state scattering of highly
  vibrationally excited {NO} at broadly tunable energies, {\it Nat. Chem.\/}
  {\bf 12},   528--534 (2020).

\bibitem{Chefdeville:Science341:06092013}
S.~Chefdeville, {\it et~al.\/}, Observation of partial wave resonances in
  low-energy {O}$_2$-{H}$_2$ inelastic collisions, {\it Science\/} {\bf 341},
  1094--1096 (2013).

\bibitem{Henson:Science338:234}
A.~B. Henson, S.~Gersten, Y.~Shagam, J.~Narevicius, E.~Narevicius, Observation
  of resonances in {P}enning ionization reactions at sub-kelvin temperatures in
  merged beams, {\it Science\/} {\bf 338},   234--238 (2012).

\bibitem{Gordon:NatChem10:119}
S.~D.~S. Gordon, {\it et~al.\/}, Quantum-state-controlled channel branching in
  cold {N}e($^3${P}$_2$) + {A}r chemi-ionization, {\it Nat. Chem.\/} {\bf 10},
   1190--1195 (2018).

\bibitem{Zhelyazkova:PRL125:263401}
V.~Zhelyazkova, F.~B.~V. Martins, J.~A. Agner, H.~Schmutz, F.~Merkt,
  Ion-molecule reactions below 1 {K}: Strong enhancement of the reaction rate
  of the ion-dipole reaction {H}e + {CH}$_{3}${F}, {\it Phys. Rev. Lett.\/}
  {\bf 125},   263401 (2020).

\bibitem{DeJongh:Science368:626}
T.~de~Jongh, {\it et~al.\/}, Imaging the onset of the resonance regime in
  low-energy {NO}-{H}e collisions, {\it Science\/} {\bf 368},   626--630
  (2020).

\bibitem{Margulis:NatCom11:3553}
B.~Margulis, J.~Narevicius, E.~Narevicius, Direct observation of a {F}eshbach
  resonance by coincidence detection of ions and electrons in {P}enning
  ionization collisions, {\it Nat. Commun.\/} {\bf 11},   3553 (2020).

\bibitem{DeJongh:NatChem14:538}
T.~de~Jongh, {\it et~al.\/}, Mapping partial wave dynamics in scattering
  resonances by rotational de-excitation collisions, {\it Nat. Chem.\/} {\bf
  14},   538--544 (2022).

\bibitem{Bohn:NJP11:055039}
J.~L. Bohn, M.~Cavagnero, C.~Ticknor, Quasi-universal dipolar scattering in
  cold and ultracold gases, {\it New J. Phys.\/} {\bf 11},   055039 (2009).

\bibitem{Cavagnero:NJP11:055040}
M.~Cavagnero, C.~Newell, Inelastic semiclassical collisions in cold dipolar
  gases, {\it New J. Phys.\/} {\bf 11},   055040 (2009).

\bibitem{Avdeenkov:PRL90:043006}
A.~V. Avdeenkov, J.~L. Bohn, Linking ultracold polar molecules, {\it Phys. Rev.
  Lett.\/} {\bf 90},   043006 (2003).

\bibitem{Frye:NJP17:045019}
M.~D. Frye, P.~S. Julienne, J.~M. Hutson, Cold atomic and molecular collisions:
  approaching the universal loss regime, {\it New J. Phys.\/} {\bf 17},
  045019 (2015).

\bibitem{Kirste:Sience338:1060}
M.~Kirste, {\it et~al.\/}, Quantum-state resolved bimolecular collisions of
  velocity-controlled {OH} with {NO} radicals, {\it Science\/} {\bf 338},
  1060--1063 (2012).

\bibitem{Langley:AO30:3459}
D.~S. Langley, M.~J. Morrell, Rainbow-enhanced forward and backward glory
  scattering, {\it Appl. Opt.\/} {\bf 30},   3459--3467 (1991).

\bibitem{Duren:CPL64:357}
R.~D{\"u}ren, H.~O. Hoppe, H.~Tischer, Observation of the orbiting backward
  peak in the differential cross section for {N}a($^2${P})-{H}g collisions,
  {\it Chem. Phys. Lett.\/} {\bf 64},   357--359 (1979).

\bibitem{Loesch:JCP99:9598}
H.~J. Loesch, F.~Stienkemeier, Evidence for the deep potential well of {L}i $+$
  {HF} from backward glory scattering, {\it J. Chem. Phys.\/} {\bf 99},
  9598--9602 (1993).

\bibitem{Bohn:PRA63:052714}
J.~L. Bohn, Inelastic collisions of ultracold polar molecules, {\it Phys. Rev.
  A\/} {\bf 63},   052714 (2001).

\bibitem{Troe:JPCA114:9762}
E.~Nikitin, J.~Troe, Mutual capture of dipolar molecules at low and very low
  energies. {I}. {A}pproximate analytical treatment, {\it J. Phys. Chem. A\/}
  {\bf 114},   9762--9767 (2010).

\bibitem{Troe:JPCA115:5027}
M.~Auzinsh, E.~Dashevskaya, I.~Litvin, E.~Nikitin, J.~Troe, Mutual capture of
  dipolar molecules at low and very low energies. {II}. {N}umerical study, {\it
  J. Phys. Chem. A\/} {\bf 115},   5027--5037 (2011).

\bibitem{Troe:JCP130:014304}
M.~Auzinsh, E.~Dashevskaya, I.~Litvin, E.~Nikitin, J.~Troe, Lambda-doublet
  specificity in the low-temperature capture of {NO} ({X} $^2{\Pi}_{1/2}$) in
  low rotational states by {C}$^+$ ions, {\it J. Chem. Phys.\/} {\bf 130},
  014304 (2009).

\bibitem{Kajita:EPJD38:315}
M.~Kajita, Loss estimation of electrostatically trapped {ND}$_3$ molecules,
  {\it Eur. Phys. J. D\/} {\bf 38},   315--322 (2006).

\bibitem{Avdeenkov:PRA86:022707}
A.~V. Avdeenkov, Dipolar collisions of ultracold polar molecules in a microwave
  field, {\it Phys. Rev. A\/} {\bf 86},   022707 (2012).

\bibitem{Karman:PRL121:163401}
T.~Karman, J.~M. Hutson, Microwave shielding of ultracold polar molecules, {\it
  Phys. Rev. Lett.\/} {\bf 121},   163401 (2018).

\bibitem{Lassabliere:PRL121:163402}
L.~Lassabli{\`e}re, G.~Qu{\'e}m{\'e}ner, Controlling the scattering length of
  ultracold dipolar molecules, {\it Phys. Rev. Lett.\/} {\bf 121},   163402
  (2018).

\bibitem{Anderegg:Science373:779}
L.~Anderegg, {\it et~al.\/}, Observation of microwave shielding of ultracold
  molecules, {\it Science\/} {\bf 373},   779--782 (2021).

\bibitem{Yan:PRL125:063401}
Z.~Z. Yan, {\it et~al.\/}, Resonant dipolar collisions of ultracold molecules
  induced by microwave dressing, {\it Phys. Rev. Lett.\/} {\bf 125},   063401
  (2020).

\bibitem{Tang:data}
G.~Tang, {\it et~al.\/}, Quantum state resolved molecular dipolar collisions
over four decades of energy \textemdash online data, Zenodo (2023); DOI: 10.5281/zenodo.7561771. 

\bibitem{Plomp:MP119:e1814437}
V.~Plomp, Z.~Gao, S.~Y.~T. van~de Meerakker, A velocity map imaging apparatus
  optimised for high-resolution crossed molecular beam experiments, {\it Mol.
  Phys.\/} {\bf 119},   e1814437 (2021).

\bibitem{Jankunas:jcp140:2014}
J.~Jankunas, B.~Bertsche, K.~Jachymski, M.~Hapka, A.~Osterwalder, Dynamics of
  gas phase {N}e* + {NH}$_3$ and {N}e* + {ND}$_3$ {P}enning ionisation at low
  temperatures, {\it J. Chem. Phys.\/} {\bf 140},   244302 (2014).

\bibitem{Crompvoets:Nat411:174}
F.~M.~H. Crompvoets, H.~L. Bethlem, R.~T. Jongma, G.~Meijer, A prototype
  storage ring for neutral molecules, {\it Nature\/} {\bf 411},   174--176
  (2001).

\bibitem{Heiner:NatPhys3:115}
C.~E. Heiner, D.~Carty, G.~Meijer, H.~L. Bethlem, A molecular synchrotron, {\it
  Nat. Phys.\/} {\bf 3},   115--118 (2007).

\bibitem{Zieger:zpc:227:2013}
P.~C. Zieger, {\it et~al.\/}, A forty-segment molecular synchrotron, {\it Z.
  Phys. Chem.\/} {\bf 227},   1605--1645 (2013).

\bibitem{Poel:prl120:2018}
A.~P.~P. van~der Poel, {\it et~al.\/}, Cold collisions in a molecular
  synchrotron, {\it Phys. Rev. Lett.\/} {\bf 120},   033402 (2018).

\bibitem{Yan:RSI84:023102}
B.~Yan, {\it et~al.\/}, A new high intensity and short-pulse molecular beam
  valve, {\it Rev. Sci. Instrum.\/} {\bf 84},   023102 (2013).

\bibitem{molpro:2008}
{M}OLPRO, version 2008.1, a package of {\it ab initio programs}, H.-J. Werner,
  P. J. Knowles, R. Lindh, F. R. Manby, M. Sch{\"u}tz, and others, see
  http://www.molpro.net.

\bibitem{Boys:MolPhys19:553}
S.~F. Boys, F.~Bernardi, The calculation of small molecular interactions by the
  differences of separate total energies. {S}ome procedures with reduced
  errors, {\it Mol. Phys.\/} {\bf 19},   553--566 (1970).

\bibitem{Ho:JCP104:1996}
T.-S. Ho, H.~Rabitz, A general method for constructing multidimensional
  molecular potential energy surfaces from {\em ab initio} calculations, {\it
  J. Chem. Phys.\/} {\bf 104},   2584--2597 (1996).

\bibitem{Wormer:JCP122:244325}
P.~E.~S. Wormer, J.~A. K{\l}os, G.~C. Groenenboom, A.~van~der Avoird, {\em Ab
  initio} computed diabatic potential energy surfaces of {OH}-{HC}l, {\it J.
  Chem. Phys.\/} {\bf 122},   244325 (2005).

\bibitem{lester:76}
W.~A. Lester, {\it The $N$ coupled-channel problem\/} (Plenum, New York, 1976),
    1--32.

\bibitem{groenenboom:10}
G.~C. Groenenboom, L.~M.~C. Janssen, {\it Tutorials in molecular reaction
  dynamics\/}, M.~Brouard, C.~Vallance, eds. (RSC, Cambridge, 2010),
  392--441.

\bibitem{Stuhl:NATURE492:396}
B.~K. Stuhl, {\it et~al.\/}, Evaporative cooling of the dipolar hydroxyl
  radical, {\it Nature\/} {\bf 492},   396--400 (2012).

\bibitem{Veldhoven:PRA66:032501}
J.~van Veldhoven, R.~T. Jongma, B.~Sartakov, W.~A. Bongers, G.~Meij\-er,
  Hyperfine structure of {ND}$_3$, {\it Phys. Rev. A\/} {\bf 66},   32501
  (2002).

\bibitem{Kukolich:PR156:83}
S.~G. Kukolich, Measurement of ammonia hyperfine structure with a two-cavity
  maser, {\it Phys. Rev.\/} {\bf 156},   83--92 (1967).

\bibitem{Meerts:MolSpec44:320}
W.~L. Meerts, A.~Dymanus, The hyperfine {$\Lambda$}-doubling spectrum of
  $^{14}${N}$^{16}${O} and $^{15}${N}$^{16}${O}, {\it J. Mol. Spectr.\/} {\bf
  44},   320--346 (1972).

\bibitem{Hoy:cjp53:1975}
A.~R. Hoy, J.~W.~C. Johns, A.~R.~W. McKellar, Stark spectroscopy with the {CO}
  laser: Dipole moments, hyperfine structure, and level crossing effects in the
  fundamental band of {NO}, {\it Can. J. Phys.\/} {\bf 53},   2029--2039
  (1975).

\bibitem{Ubachs:JMS115:88}
W.~Ubachs, G.~Meijer, J.~J. {ter Meulen}, A.~Dymanus, High-resolution
  spectroscopy on the $c\,^1{\Pi} \leftarrow a\, ^1{\Delta}$ transition in
  {NH}, {\it J. Mol. Spectrosc.\/} {\bf 115},   88--104 (1986).

\bibitem{Hain:JPCA101:7674}
T.~D. Hain, M.~A. Weibel, K.~M. Backstrand, T.~J. Curtiss, Rotational state
  selection and orientation of {OH} and {OD} radicals by electric hexapole
  beam-focusing, {\it J. Phys. Chem. A\/} {\bf 101},   7674--7683 (1997).

\bibitem{Weibel:jcp108:1998}
M.~A. Weibel, T.~D. Hain, T.~J. Curtiss, Hexapole-selected supersonic beams of
  reactive radicals: {CF}$_3$, {S}i{F}$_3$, {SH}, {CH}, and {C}$_2${H}, {\it J.
  Chem. Phys.\/} {\bf 108},   3134--3141 (1998).

\bibitem{Drabbels:cpl200:108}
M.~Drabbels, S.~Stolte, G.~Meijer, Production of an intense pulsed beam of
  oriented metastable {CO} $a\,^3{\Pi}$, {\it Chem. Phys. Lett.\/} {\bf 200},
  108--112 (1992).

\bibitem{Eyles:NatChem3:597}
C.~J. Eyles, {\it et~al.\/}, Interference structures in the differential cross
  sections for inelastic scattering of {NO} by {A}r, {\it Nat. Chem.\/} {\bf
  3},   597--602 (2011).

\bibitem{Besemer:NatChem14:664}
M.~Besemer, {\it et~al.\/}, Glory scattering in deeply inelastic molecular
  collisions, {\it Nat. Chem.\/} {\bf 14},   664--669 (2022).

\bibitem{Adam:PhysRep356:229}
J.~A. Adam, The mathematical physics of rainbows and glories, {\it Phys.
  Rep.\/} {\bf 356},   229--365 (2002).

\bibitem{korsch:JPBAMP16:793}
H.~Korsch, K.~E. Thylwe, Orbiting and rainbow structures in low-energy elastic
  differential cross sections, {\it J. Phys. B: At. Mol. Phys.\/} {\bf 16},
  793--815 (1983).

\bibitem{friedrich:13}
H.~Friedrich, {\it Scattering Theory\/} (Springer, Berlin, 2013).
\end{thebibliography}
\bibliographystyle{Science}

{\bf Acknowledgement}
We thank Daan Snoeken for help in developig the NO-ND$_3$ PES, Youp Caris for assistance during the ICS measurements, and Zhi Gao for assistance during the high energy DCS measurements. We thank Henrik Haak and Gerard Meijer of the Fritz-Haber-Institut der Max-Planck-Gesellschaft for providing the curved hexapole. We thank Frank Stienkemeier for fruitful discussions on earlier work on backward glories. We thank Niek Janssen and Andr\'e van Roij for expert technical support.
{\bf Funding:}  This work is part of the research program of the Netherlands Organization for Scientific Research (NWO) through an ENW-M and a VICI grant. S.Y.T.v.d.M. acknowledges support from the European Research Council (ERC) under the European Union's Horizon 2020 Research and Innovation Program (Grant Agreement No. 817947 FICOMOL). G.T. acknowledges support from the China Scholarship Council under grant 201606340077. {\bf Author contributions:} The experiments were performed and analyzed by G.T.. The theoretical calculations were performed by M.B., T.K., G.C.G. and A.v.d.A.. Simulations on beam overlap were performed by S.K. The project was conceived and supervised by S.Y.T.v.d.M.. The main manuscript was written by S.Y.T.v.d.M.. The Supplement was written by G.T., M.B. and S.Y.T.v.d.M.. All authors were involved in the interpretation of the data, discussed the results, and commented on the manuscript. {\bf Competing interests:} None declared. {\bf Data and materials availability:} All  data needed to evaluate the conclusions in the paper are present in the paper or the Supplementary Materials, as well as online at DOI: 10.5281/zenodo.7561771 \cite{Tang:data}.
\newpage

{\bf Supplementary Materials}\\
Supplementary text\\
Figures S1 - S35\\
Tables S1- S4\\
References 55-81  

\begin{table}[!tp]
	\centering
	\caption{Prediction for the collision energy (upper entry, in cm$^{-1}$) and maximal cross section (lower entry, in \AA$^2$) of the LM in the ICS for collisions between various combinations of the molecules NH$_3$, ND$_3$, NO ($X\,^2\Pi_{1/2}$), NH ($a\,^1\Delta$), OH ($X\,^2\Pi_{3/2}$) and CO ($a^3\Pi$). For each molecule, the rotational state with maximal Stark shift is assumed (see SM for details).}
	\begin{center}
		\renewcommand\arraystretch{1.2}
		\begin{tabularx}{16cm}{p{0.6cm}<{\centering}X<{\centering}X<{\centering}X<{\centering}X<{\centering}X<{\centering}X<{\centering}}
			\hline
			\hline
	                & ND$_3$ & NH$_3$ & NO & NH & OH & CO        \\
			\hline
			ND$_3$ & 0.28 & 2.20 & 0.17 & 0.14 & 0.28 &  0.17     \\
			      & 4457 & 1101 & 1061 &  8531 & 5319 &  5748     \\
			NH$_3$& 2.20 & 4.12 & 2.09 & 2.06 & 2.20 &  2.09     \\
			      & 1101 & 714  & 195  & 1386 & 1333 &  1070     \\
			NO & 0.17 & 2.09 & 0.06 & 0.03 & 0.18 & 0.06  \\
			         & 1061 & 195  & 355  &  3963 & 1254 & 1886  \\			
			NH & 0.14 & 2.06 & 0.03 & 2.01$\times$10$^{-5}$ & 0.14 &  0.03         \\
			  & 8531 & 1386 & 3963 &  3.71$\times$10$^6$    & 1.00$\times$10$^4$ & 2.04$\times$10$^4$   \\
			OH & 0.28 & 2.20 & 0.18 & 0.14 & 0.29 &  0.18       \\
			  & 5319 & 1333 & 1254 & 1.00$\times$10$^4$ & 6351 &  6795 \\
			CO& 0.17 & 2.09 & 0.06 & 0.03 & 0.18 &  0.07       \\
			  & 5748 & 1070 & 1886 & 2.04$\times$10$^4$ & 6795 & 1.00$\times$10$^4$ \\
			\hline
			\hline
         \end{tabularx}
	\end{center}
	\label{tab:Esp_Ecol}
\end{table}

\end{document}